\documentclass{ttp_DSL2006}
\usepackage[dvipsone]{graphicx}
\usepackage[intlimits]{amsmath}
\usepackage{amssymb}
\usepackage{exscale}
\begin{document}
\title{Theoretical Investigation of the Magnetic Order in FeAs}
\author{Dobysheva Lyudmila V.
$^{\rm{a}}$ and Arzhnikov Anatoly K.$^{\rm{b}}$}
\institute{Phys.-Techn. Institute UrBr Russian Ac. Sci.,.
426001, Izhevsk, Kirova str., 132, RUSSIA}
\maketitle
\vspace{-3mm}
\sffamily
\begin{center}
$^{a}$lyu@otf.pti.udm.ru, $^{ b}$arzhnikov@otf.pti.udm.ru
\end{center}

\vspace{2mm} \hspace{-7.7mm} \normalsize \textbf{Keywords:}
iron monoarsenid, FeAs, first-principles calculation, 
magnetic state, collinear state, elliptic spin spiral, 
magnetic anisotropy.\\

\vspace{-2mm} \hspace{-7.7mm}
\rmfamily

\noindent \textbf{Abstract.} 
The magnetic structure of the iron monoarsenide FeAs is studied using
first-principles calculations. We consider the collinear and
non-collinear (spin-spiral wave) magnetic ordering and magnetic
anisotropy.  It is analitically shown that a magnetic triaxial anisotropy
results in a sum of two spin-spiral waves with opposite directions of
wave vectors and different spin amplitudes, so that the magnetic moments
in two perpendicular directions do not equal each other.

\vspace{-2mm}
\section{Introduction}
\vspace {-3mm}

\hspace{4.5mm}
The discovery in 2008 of a new class of superconductors, namely, layered
compounds on the basis of iron \cite{Kamihara} attracted a great
interest. This broke the monopoly of cuprates in the
high-temperature-superconductivity (HTSC) physics and aroused an
expectance for the progress in theoretical understanding of the HTSC.
The new class of superconductors demonstrated both common and different
features in comparison to cuprates \cite{Mazin}. One of the common points is
a presence of a strong magnetisation. The majority of researchers
believe that, in these superconductors, magnetic fluctuations are
responsible for the electron pairing as in the case of cuprates. That is
why numerous works, including the presented one, are devoted to
investigation of the magnetic properties.

The FeAs attracts large interest due to extraordinary properties of its
close relatives such as LaFeAsO, BaFe2As2, and NaFeAs, which are
attributed to the presence of FeAs planes with the same local structure
as in the iron monoarsenide. The experimental work conducted in 2011
\cite{Rodriguez} using a polarised neutrons revealed a row of
peculiarities which are important for theory. In particular, direct
measurements of spin-spiral waves that are the ground state of the system
were conducted. However, the authors interpretation in that work based
on the fitting of the parameters of a Heisenberg Hamiltonian is
misleading, as d-electrons which form the magnetic moment are delocalised, and correlations are weak
\cite{Mazin}.

In connexion  with this, here, a first-principles calculation of
magnetic properties of FeAs crystal is conducted and an interpretation
of the anisotropy of the magnetic moment of spin-spiral wave is given in
the frames of a model Hamiltonian.

\section{First-principles calculation of the electron structure}
\vspace {-3mm}

The crystal structure of FeAs is well known from experiments, for example, 
\cite{Selte}.
It forms the orthorhombic (Pnma) MnP-type crystal (Fig.~1). 
\begin{figure}[!ht]\includegraphics*[width=8.8cm]{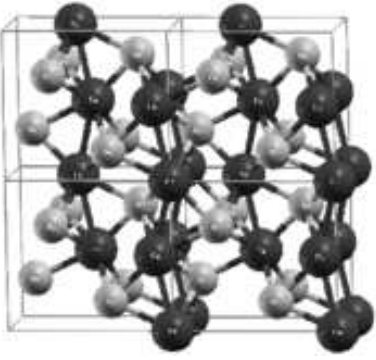}
\caption{2x2 unit cell of FeAs.} \label{cellFeAs}\end{figure}
 
The calculations are conducted with a full-potential linearized
augmented waves (FP-LAPW) method realized in the package WIEN2k \cite{WIEN2k}. In
this method the unit cell is divided into two parts: the nonoverlapping
atomic spheres and the interstitial region. In the latter the
wave functions are expanded into plane waves; inside the spheres
the plane waves are augmented by an atomic-like spherical harmonics
expansion with radial wave functions.  The exchange - correlation
potential is calculated using the "local density approximation"
(LDA) \cite{LDA} and the "generalized gradient approximation" (GGA)
\cite{GGA}. All parameters of the calculation scheme are
chosen in a standard for the method way. The basic results are tested
for numerical convergence with respect to the calculation parameters,
they do not also depend on the approximation of the exchange-correlation
potential: GGA \cite{GGA} or LDA \cite{LDA}.

With the experimental lattice parameters \cite{Selte}
a = 0.60278 nm, b = 0.54420 nm, and c = 0.33727 nm, the precise positions of 
atoms in the unit cell that correspond to the minimum energy are found 
through optimisation of internal parameters in the GGA potential. The 
optimized positions are as follows: \\
Fe1 = (0.0, 0.0, 0.0), 
Fe2 = (0.39895, 0.99961, 0.5), \\
Fe3 = (0.89895, 0.5, 0.0), 
Fe4 = (0.5, 0.49961, 0.5), \\
As1 = (0.62404, 0.19863, 0.0), 
As2 = (0.77491, 0.80098, 0.5), \\
As3 = (0.27491, 0.69863, 0.0), 
As4 = (0.12404, 0.30098, 0.5), \\
which are close to the experimentally found positions.

Further, using the experimental lattice parameters and the optimised 
atomic positions, considered are four magnetic states 
with collinear magnetic moments. In
one of them all Fe local magnetic moments have the same direction 
(ferromagnetic FM order), the other three have both positive and 
negative moments with zero overal magnetization and are denominated as 
AFM1 (Fe1 up, Fe2 up, Fe3 down, Fe4 down), 
AFM2 (Fe1 up, Fe2 down, Fe3 up, Fe4 down), and
AFM3 (Fe1 up, Fe2 down, Fe3 down, Fe4 up).
Different collinear solutions (FM and AFMs) are obtained using different 
starting potentials.

The AFM solutions are slightly lower in energy than the FM state (Table 1).
\begin{table*}[!h]\vspace {-5mm}\caption{The energy $E$ and atomic magnetic moments $M$ in four collinear magnetic states considered in FeAs}
\begin{tabular}{|c|cc|}\hline 
magnetic state&$E$(Ry)& $M$\\
\hline 
FM       &-28271.25146&0.62 \\
AFM1 uudd&-28271.25916&1.00 \\
AFM2 udud&-28271.26015&0.99 \\
AFM3 uddu&-28271.26471&1.23 \\
\hline \end{tabular}\label{Table1} \end{table*}
The state AFM3 is lowest by energy, the FM state being higher by 13 mRy.
A calculation with account of the spin-orbit interaction 
is conducted and the magnetocrystalline anisotropy is obtained. 
The magnetocrystalline anisotropy obtained as a difference of 
total energy between the states with magnetisation along the directions 
specified is equal to 
$E_{010}-E_{001}=1 \mu$Ry and 
$E_{100}-E_{001}=36 \mu$Ry.

The optimization of the lattice parameters in the LDA exchange-correlation 
potential confirms a well-known tendency: the GGA potential allows to obtain 
structural parameters closer to experiment, whereas the LDA potential 
gives the parameters further from the experimental ones but describes better
than the GGA potential magnetic moment which is $M_{LDA}= 0.65 \mu_B$ 
for the AFM3 state (compare to experiment in \cite{Rodriguez} 
$M=0.58\pm 0.06 \mu_B$).

The possibility of a few magnetic states of a system means that an 
intermediate state is the state with a minimum energy. We
conducted a calculation of energy of spin-spiral wave depending on its
wave vector $q$. The calculation is performed with a
non-collinear-magnetic version of WIEN2k package \cite{NCM1,NCM2}.

$E(q)$ curves obtained show that the most energetically favourable 
state AFM3 has a minimum energy at $q=0$ (the directions 100 and 001 of
q were considered). 
\begin{figure}[!h]\includegraphics*[width=8.8cm]{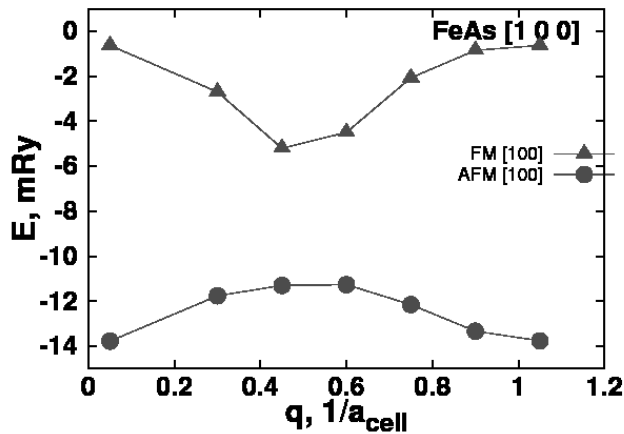}
\caption{Energy dependence of the spin-spiral wave on the q wave vector in FeAs.} 
\label{E_q-FeAs}\end{figure}
In the FM state, the SSW with the wave vector $q=0.5 a^{-1}$ in 100 
direction possesses the minimum energy (here $a$ is the lattice
parameter), the SSW with the 001 direction of $q$ having a minimum at
$q=0$. The difference between the collinear and the SSW solution at
minimum energy is about 5 mRy/cell, see fig.2.

We have also studied the solutions at the lattice parameters obtained 
in the LDA of exchange-correlation potential after minimization by
energy. Qualitatively the results repeat, though the numerical
differences between the states are smaller.

\section{Anisotropy}
\vspace{-3mm}
The calculations of the state with minimum energy, that is the AFM3
state, show that the system has a triaxial anisotropy, the energy of the
state with magnetisation perpendicular to the iron planes (that is $M ||$
100) being significantly higher than in cases with $M ||$ 001 or 010. So, the
magnetic moment should lie in the plane of iron atoms, which is
confirmed  experimentally in \cite{Rodriguez}. At the same time, there
is a slight magnetic anisotropy in the plane between the directions 010
and 001 (the difference of 1 $\mu$Ry is on the level of calculational errors).
Therefore we assume that magnetic anisotropy of the FeAs compound can be
described by an angle between an easy axis $x$ in the iron plane which we 
denote here as $xy$, or by the projection of the magnetic moment to the x axis:
\begin{eqnarray}
E_{anis}=-\sum_{i}\sum_{i} 4 D(S_i^{(x)})^2
\end{eqnarray}
Adding this contributon to the Hubbard Hamiltonian, we obtain:
\begin{eqnarray}
H=\frac{1}{N}\sum_{k,\sigma}\epsilon_k^0 c^+_{k \sigma} c^{ }_{k \sigma} -
\frac{U}{2N}\sum_{j} 
c^+_{j \uparrow} c^{ }_{j \uparrow} c^+_{j \downarrow} c^{ }_{j \downarrow} -
\sum_{j} D 
(c^+_{j \uparrow} c^{ }_{j \downarrow} + c^+_{j \downarrow} c^{ }_{j \uparrow})^2
\end{eqnarray}
Here we use the equality
\begin{eqnarray}
2 S^x_j = (S^+_j + S^-_j) = 
(c^+_{j \uparrow} c^{ }_{j \downarrow} + c^+_{j \downarrow} c^{ }_{j \uparrow})
\end{eqnarray}

The simplest approximation of this Hamiltonian over the class of the
wave functions of the spin-spiral waves is a mean field approximation
(MFA). Then the Hamiltonian of the system with anisotropy can be
rewritten as:
\begin{eqnarray}
H= & (U+3D)(M_1^2+M_2^2) +U n_{el}/2+
1/N \sum_{k,\sigma}\epsilon_k^0 c^+_{k \sigma} c^{ }_{k \sigma} \\
& + \left [ -(U+2D)M_1-D M_2 \right ] \frac{1}{N}\sum_Q (S^+_Q + S^-_Q) \nonumber \\
& + \left [ -(U+2D)M_2-D M_1 \right ] \frac{1}{N}\sum_Q (S^+_{-Q} + S^-_{-Q}) \nonumber
\end{eqnarray}
where $S^+_Q=\sum_k c^+_{k \uparrow} c_{k+Q \downarrow}$,
$S^-_Q=\sum_k c^+_{k \uparrow} c_{k-Q \downarrow}$,
$\langle S^+_Q\rangle=\langle S^-_Q\rangle= M_1$,
$\langle S^+_{-Q}\rangle =\langle S^-_{-Q}\rangle= M_2.$

Solution of the system is a superposition of two spin-spiral
wave of opposite directions with a total magnetisation: 
\begin{eqnarray}
\overrightarrow{M} =
\overrightarrow{e_x} (M_1+M_2) \cos \overrightarrow{Q} \overrightarrow{R_j} + 
\overrightarrow{e_y} (M_1-M_2) \sin \overrightarrow{Q} \overrightarrow{R_j}
\end{eqnarray}

Different projections of the magnetization on the axis x and y were 
actually revealed in experiment \cite{Rodriguez}.

\vspace{-3mm}
\section{Summary}
\vspace{-3mm}
\noindent 
Using an ab-initio method of the electron structure calculation (FP LAPW
that is realized in WIEN2k package), we have studied the magnetic
structure of FeAs, both GGA and LDA potentials being used in
calculations. After finding the optimum atomic positions in the unit
cell, we have considered a few magnetic configurations with a collinear
structure. There are a ferromagnetic FM and three kinds of
antiferromagnetic AFM structures found. The AFM structures have lower
total energies than the FM one. For all four structures, the
calculations with spin-orbit term included have been conducted and the
magnetic anisotropy has been studied. 

Using a package version for the noncollinear magnetism, the dependence
of the total energy on the wave vector q has been obtained. 

With a model Hamiltonian, it has been shown that triaxial magnetic
anisotropy results in a spin-spiral waves with a corresponding
difference in the spin amplitudes, so that the  magnetic moments in two
perpendicular directions  do not equal each other, and the spin spiral
becomes elliptic.

Support by RFBR (grant N 09-02-00461) is acknowledged.

\end{document}